\documentclass[aps,pre,twocolumn,superscriptaddress,amsmath,amssymb]{revtex4-1}
\usepackage{graphicx}  
\usepackage{dcolumn}   
\usepackage{bm}        
\usepackage{verbatim}   
\usepackage[utf8]{inputenc}
\usepackage{tabularx}
\usepackage{xspace}      
\usepackage{dsfont}
\usepackage[decimalsymbol=.]{siunitx}
\usepackage{xfrac}
\usepackage{lineno}      
\usepackage{color}
\usepackage{caption}
\usepackage{stfloats}

\makeatletter
\def\oversortoftilde#1{\mathop{\vbox{\m@th\ialign{##\crcr\noalign{\kern3\p@}%
      \sortoftildefill\crcr\noalign{\kern3\p@\nointerlineskip}%
      $\hfil\displaystyle{#1}\hfil$\crcr}}}\limits}

\def\sortoftildefill{$\m@th \setbox\z@\hbox{$\braceld$}%
  \braceld\leaders\vrule \@height\ht\z@ \@depth\z@\hfill\braceru$}

\makeatother

\begin{document}
\title{How coherent structures dominate the residence time in a bubble wake: an experimental example}
\author{A. v.Kameke}
\address{
   Institute of Multiphase Flows, Hamburg University of Technology\\
   Hamburg, Germany
   }
\author{S. Kastens}
\address{
   Institute of Multiphase Flows, Hamburg University of Technology\\
   Hamburg, Germany
   }
\author{S. Rüttinger}
\address{
   Institute of Multiphase Flows, Hamburg University of Technology\\
   Hamburg, Germany
   }
\author{S. Herres-Pawlis}
\address{
  Institute for Inorganic Chemistry, RWTH Aachen University, Germany
   }
   \author{M. Schlüter}
\address{
   Institute of Multiphase Flows, Hamburg University of Technology\\
   Hamburg, Germany
   }

\date{\today}
\begin{abstract}
Mixing timescales and residence times in reactive multiphase flows can be essential for product selectivity. For instance when a gas species is consumed e.g. by a competitive consecutive reaction with moderate reaction kinetics where reaction timescales are comparable to relevant mixing timescales. To point out the importance of the details of the fluid flow, we analyze experimental velocity data from a Taylor bubble wake by means of Lagrangian methods. By adjusting the channel diameter in which the Taylor bubble rises, and thus the rise velocity, we obtain three different wake regimes. Remarkably the normalized residence times of passive particles advected in the wake velocity field show a peak for intermediate rise velocities. This fact seems unintuitive at first glance because one expects a faster removal of passive tracers for a faster overall flow rate. However, the details of the flow topology analyzed using Finite Time Lyapunov Exponent (FTLE) fields and Lagrangian Coherent Structures (LCS) reveal the existence of a very coherent vortical pattern in the bubble wake which explains the long residence times. The increased residence times within the vortical structure and the close bubble interface acting as a constant gas species source could enhance side product generation of a hypothetical competitive consecutive reaction, where the first reaction with the gas species forms the desired product and the second the side product. 
\end{abstract}

\maketitle

\section{Introduction}

Multiphase flows are essential for many chemical and biochemical industrial applications like oxidation reactions, hydrogenation reactions or fermentations, where a transferred gas species is consumed by a reaction or cells in a continuous liquid phase. In all these processes the generation of  undesired side species is reported especially if the operation conditions like flow regime and mass transfer rates in the reactor are not set adequately. Therefore, empirical correlations are used\cite{Schlueter1992,Schulzke1998,Fitzer1995} to estimate the yield and selectivity of the products and side products, which are generally used to describe the efficiency of production plants in relation to the energy demand\cite{Baldyga1999}. However, yield and selectivity of a chemical reaction highly depend on the local transport processes and reaction kinetics which are coupled in a complex way and can influence each other\cite{Paul2018}. This makes it desirable to study the problem in simplified experiments which allow to achieve the high spatio-temporal resolution needed for a detailed analysis of the local processes.\\ Kastens et al.\cite{Kastens2017} have suggested to use elongated bubbles in vertical round channels, so called Taylor bubbles, in a counter current flow to ensure accurate local measurements of the fluid flow and the concentration field around those bubbles. The advantages are the small required fluid volume, the high predictability of bubble rise trajectories, easy optical access, reduced complexity and the high reproduceability of the experiments. \\
Another advantage of Taylor bubbles is their volume independent rising velocity, which can be predicted by the dimensionless Eötvös number $Eo_D$ of the fluidic system\cite{Hayashi2011} 

\begin{equation}
Eo_{D}=\dfrac{(\rho_{L}-\rho_{G})\,g\,D^{2}}{\sigma} 
\end{equation}

where $\sigma$ is the interfacial tension, $\rho_L$ and $\rho_G$ are the densities of the liquid and gaseous phases, $g$ is the magnitude of the gravitational acceleration, and $D$ is the hydraulic diameter of a round channel. Based on the work of Hayashi et al.\cite{Hayashi2011}, the critical Eötvös number is $Eo_{crit.}=4$, where the buoyancy dominates the interfacial tension and the Taylor bubble rises\cite{Bretherton1961,White1962}. For a water-gas system this leads to an inner diameter of the round channel of $D=5.4 \, mm$. At low Eötvös numbers, as long as they are elongated in the pipe, no shape oscillations occur, so that the bubble is self-centering within the round channel and during the dissolution process only the length of the bubbles is decreasing\cite{Kastens2015}. By varying the inner diameter $D$ of the round channel a variation of the rising velocity $v_b$ and, therefore the wake structure is observed\cite{Kastens2015}. 

In this study we take advantage of this well-defined adjustability of rising velocities and shapes of the Taylor bubbles and create three characteristic and reproducible wake structures. We analyze the velocity fields in the bubble wakes using particle image velocimetry (PIV) and numerically advect passive particles in the fields derived. This procedure forms the basis for a detailed lagrangian analysis of the flow structure applying the actual theoretical concepts by using an adapted open access MATLAB toolbox\cite{Onu2015, Haller2015}. The Lagrangian analysis reveals the underlying flow topology that orders the particle transport. Especially for the intermediate wake regime a persistent vortical structure appears to very dominantly influence the mixing of tracers of different fates and origins. An analysis of the residence times of the passive particles reveals a peak for the intermediate wake regime. The  implications that the details of the  lagrangian topology have for a hypothetical chemical reaction are discussed.\\

\section{Methods}
\label{sec:methodology}

\subsection{Experimental Settings}
\begin{center}
\begin{figure}[p]
	\includegraphics[width=\linewidth]{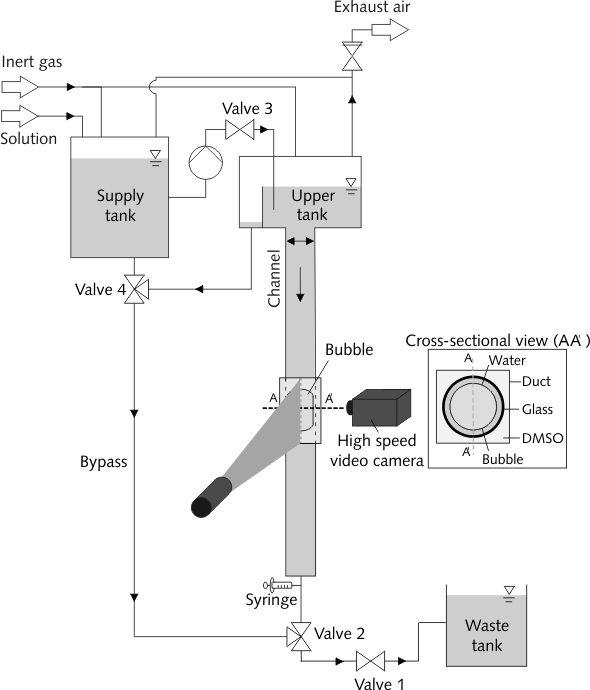}
	\captionof{figure}{Schematic experimental setup ($get the permission of reproduction from CET$)}\label{fig:-1}
	\end{figure}
\end{center}

\begin{figure*}[t]
	\includegraphics[width=\textwidth]{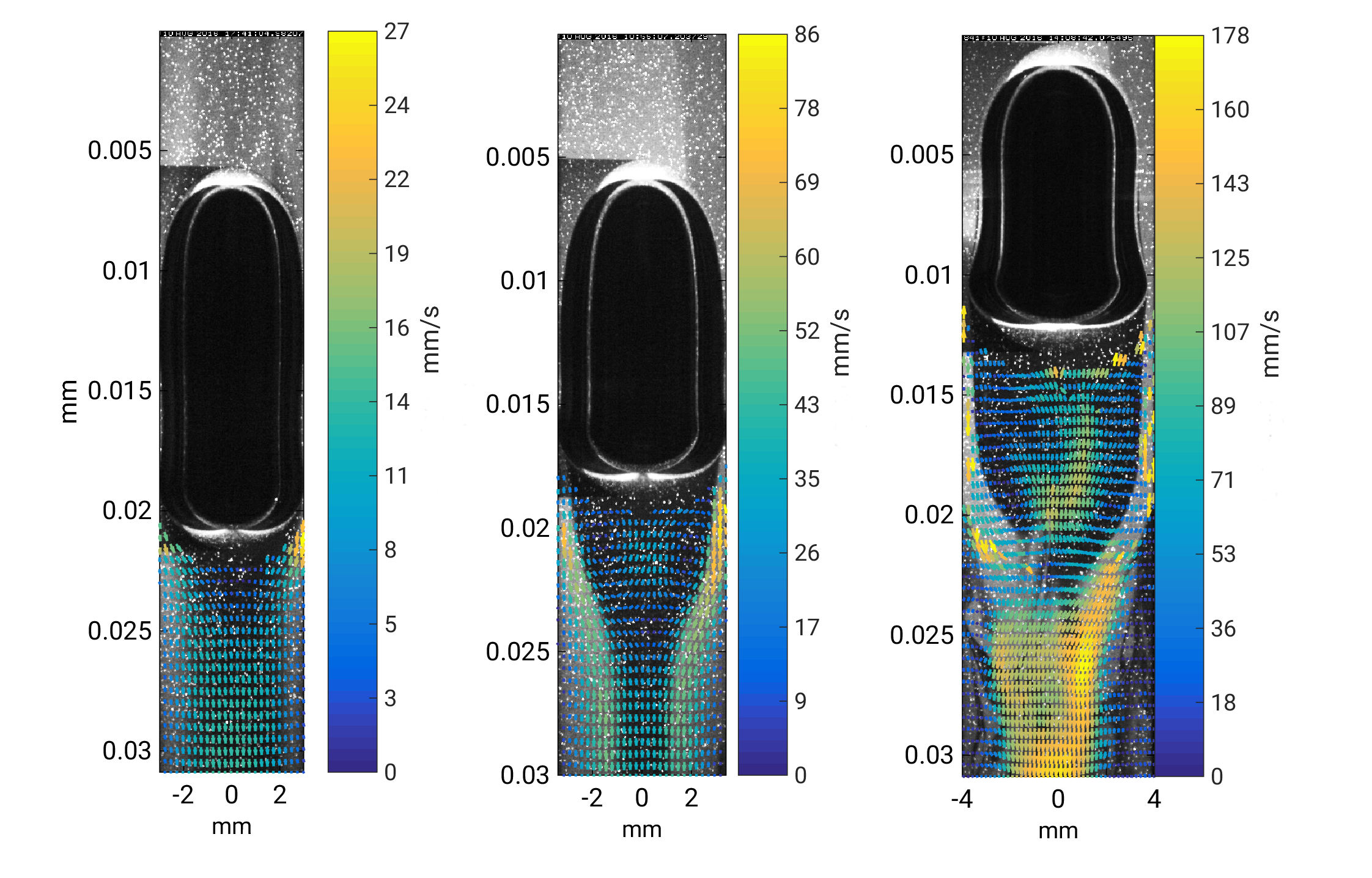}
	\captionof{figure}{The instantaneous velocity fields from PIV measurements and simultaneous LIF for three different flow regimes. Left to right: $D = 6\,mm$, $D = 7\,mm$, $D = 8\,mm$ round channel diameter. }\label{fig:0}
\end{figure*}

The experimental set-up\cite{Kastens2017} consists of a vertical round glass channel ($L=300 \,mm$) as test section connected to an upper tank as illustrated in Fig.\ref{fig:-1}. The upper tank includes an overflow device to keep the liquid level and therefore the pressure within the vertical round channel constant. The liquid down flow rate within the round channel can be adjusted by using control valve 1 to keep the bubble fixed at $360\,mm$ below the water surface in the upper tank. As test sections, three circular pipes, $D = 6, 7$ and $8\, mm$ (manufactured by Glastechnik Kirste KG) have been used to adjust the bubble rising velocity and thus obtain three different wake flow regimes at different flow Reynolds numbers $Re = 36$, $Re = 156$ and $Re = 303$. These were calculated using the corresponding velocities of the rising Taylor bubble in the absence of counter flow: $v_{rise} = 6 \, mm/s$, $v_{rise} = 22 \, mm/s$ and $v_{rise} = 38 \, mm/s$, see Fig.\ref{fig:0}. The bubble of the reactive gas, here $CO_{2}$, is injected by a precise gas-tight syringe (Hamilton 1001) through a septum into the lower part of the round channel. The round channel is surrounded by a square duct, made of borosilicate glass. The gap is filled with a solution containing water and dimethyl sulfoxide (DMSO, Sigma Aldrich\textsuperscript{\textregistered}) for refractive index matching (Fig.\ref{fig:-1} right, cross-sectional view). For an optical access into the round channel with minimized optical disturbance, the concentration and temperature of the DMSO solution was kept at $97 \, wt\%$  and $T_{DMSO} = 298 $±$ 1.0\,K$, respectively, to obtain the same refractive index as borosilicate glass ($n = 1.473$). The apparatus is connected via valve 2 with a supply tank to provide the solutions for the experiments.
Here we adjust a volume flow rate of the down flowing liquid phase as such that the absolute position of the bubble stays constant. This technique is especially suitable for measurements with longer integration times needed for the residence time analysis. For the visualization and quantitative analysis of the wake structures behind Taylor bubbles, $CO_{2}$ ($99.995\,vol\,\%$ purity, Westfalen AG) is injected as gaseous phase into deionized water as continuous liquid phase. The measurement of the local flow field and concentration wake structure has been performed simultaneously, by means of particle image velocimetry (PIV) and laser-induced fluorescence (LIF). Therefore, tracer particles (microParticles GmbH, PS-FluoRot-3.0, $d_{p}= 3.16 \,\micro m$) were added to a fluorescein sodium saltsolution (Sigma Aldrich  \textsuperscript{\textregistered}, $10^{-5} mol L^{-1}$).

Because fluorescein shows a decrease in fluorescence intensity as a function of the pH value of the liquid, the dissociated carbon dioxide concentration can be visualized to identify  time depending flow structures in unsteady wakes vortices. As a light source for excitation a Nd:YLF laser (Darwin Duo $527-100\,M$, $\lambda=527\,nm$, pulse width $<210\,ns$, pulse repetition rate $600\,Hz$, Quantronix) is applied. To illuminate a planar area around the hydrodynamically fixed or rising Taylor bubble, the laser beam is widened with light sheet optics (ILA 5150 GmbH). The emitted light from the fluorescence dye and the tracer particles are recorded with a PCO Dimax HS2 at a rate of $600\,fps$ ($\delta t = 1/600\,s$) perpendicular to the laser sheet with equipped bandpass filter to protect the camera from direct laser light (ILA\_5150 GmbH, $590\,nm $±$ 2\,nm$, half-power bandwidth $20\,nm $±$ 2\,nm$, transmission $> 84\,\%$). The liquid temperature $T_L$ was set constant at $T_L=298$±$0.5\,K$ using a Thermostat . 

To derive the two-dimensional velocity fields from the particle images we employ the PIV software PIVview 2C 3.63, PIVTec GmbH, with parameters as stated in table \ref{table-PIV}. The resulting accuracy of the PIV was assessed in the case of the PIV measurement of the lowest Reynolds number via the signal to noise ratio (SNR) which has a gaussian distribution with a mean value of $40$. Also a correlation coefficient of about $0.9$ and  the peak1/peak 2 ratio higher than $10$ everywhere apart from areas very close to the channel boundaries assure valid measurements. Further we directly estimate the error in the velocity measurement for the lowest Reynolds number from PIV analysis in the circular channel around $1/3$ bubble lengths ahead of the bubble where the velocity should be laminar and constant in time. The error of the instantaneous velocity was estimated as the temporal standard deviation of the measured vertical velocity at a point lying on the centerline ahead of the bubble. It was found to be lower than to lower than $2-3\,\%$.

\begin{table}[p]
	\centering
	\scriptsize
	\renewcommand{\arraystretch}{1.4}
	\caption{Details of the PIV System, conditions of the experiments and parameters of post processing}
	\label{table-PIV}
	\begin{tabular}{p{2cm}|p{5.5cm} r}
		\hline
		PIV Type	& 2D2C (two dimensional, two velocity components)  \\ \hline
		Laser		& Darwin Duo 527-100M (Nd-YLF laser, $\lambda=527 \,nm$, pulse energy up to $30 mJ$), Quantronix (Continuum) \\ \hline
		Sheet optics& widend angle 15°, sheet thickness approx. $<1\,mm$, ILA\_5150 GmbH\\ \hline
		Camera, Lens \& Filter	&  
		\begin{tabular}{l p{5cm}}
			$\bullet$ &  pco.dimax HS2, PCO AG \\
			$\bullet$ &  Makro-Planar T* 2/50 $\,mm$ ZE, Carl Zeiss AG \\
			$\bullet$ &  band-pass filter BrightLine HC $590\pm20\,nm$, Semrock  \\
		\end{tabular} \\ \hline
		
		Seeding	& 
		\begin{tabular}{l p{5cm}}
			$\bullet$ &  Polystyrol fluorescent particles, size 3.16 $\,\mu$m, Microparticles GmbH  \\
			$\bullet$ &  approximately 0.1 mL of suspension per 1 L of liquid \\
		\end{tabular} \\ \hline
		image number	& approximately $1000$  \\\hline
		PIV Software & PIVview 2C 3.63, PIVTec GmbH)\\\hline
		PIV processing parameters & 
		\begin{tabular}{l p{5cm}}
			$\bullet$ &  Number of passes: 3 \\
			$\bullet$ &  Final interrogation area size: 32 x 32 px\\
			$\bullet$ &  Final vector spacing: 16 x 16 px\\
			$\bullet$ &  Image deformation: bicubic spline\\
			$\bullet$ &  Image interpolation: logarithmic, bicubic spline\\
			$\bullet$ &  Validation: floating median test\\
			$\bullet$ &  Peak interpolation: 2 x 3 points, logarithmic\\
		\end{tabular}
		
		\\\hline
		Spatial resolution 	& 
		\begin{tabular}{l p{5.5cm}}
			$\bullet$ &  Image scale: 35 $\mu$m/px \\
			$\bullet$ &  Field of view: 11.49 x 30.0 mm2\\
			$\bullet$ &  Vector spacing: 0.56mm\\
			$\bullet$ &  Interrogation area size: 0.46 mm\\
		\end{tabular}
		\\\hline
		Temporal resolution	& $\Delta$t=1667 $\mu$s, $\Delta t=5 \,\times \,1667  \,\mu$s for $D = 6\,mm$
		
		\\\hline
	\end{tabular}	
\end{table}	

\subsection{Lagrangian Analysis}
In order to highlight the flow topology in the wake of the Taylor bubble we calculate local Lagrangian quantities from the measured two-dimensional velocity field such as local mean residence time distributions, Lagrangian coherent structures (LCS) and the finite-time Lyapunov exponent fields (FTLE-fields). LCS are defined to be the most repelling, attracting or shearing material lines of the tracer field in finite-time\cite{Onu2015, Haller2000, Haller2015}. Here, we concentrate on the so called hyperbolic LCS which correspond to the most attracting and repelling material lines and divide the flow into domains with regard to different tracer fates after a time $\tau$. In particular, we use the Matlab toolbox \textit{LCS Tool} which extracts LCS as null geodesics of appropriate Lorentzian metrics\cite{Onu2015}.
Mathematically the forward FTLE is defined as $\Lambda_{+}(x_{0} , t_{0} , \tau ) =  \frac{1} {\tau}  ln(  \sqrt{ \lambda_{1}(x_{0} , t_{0}) }  )$, where $\lambda_{1}$ is the largest Eigenwert of the Cauchy-Green strain tensor while the hyperbolic LCS is calculated from the Cauchy-Green strain tensor taking into account the information from the Eigenvectors as well as detailed in\cite{Onu2015}. The conclusions drawn from the FTLE-ridges (lines of highest stretching  rate in the FTLE-fields) and the LCS largely coincide for our analysis. Historically  FTLE-ridge detection is the precedent method to extract the ordering manifolds of time dependent flows but it was recently shown that newer methods like the LCS are more correct\cite{Haller2015}. Nevertheless here we find the calculation of the  FTLE-field especially useful as a measure of mixing rate since it provides stretching (contraction) information on every point regardless wether or not it lies on a distinguished manifold.

We extract repelling and attracting LCS and FTLE-fields at a time $t$ by integrating a grid of $100\times100$ equidistant distributed particles in forward and in backward time till times $t +\tau$ and $t -\tau$, respectively. The integration time was set to $\tau = 120 \,\delta t \approx 0,2\,s$. It was chosen somewhat arbitrary as a trade off between the existence of very clear and elongated FTLE and LCS structures and the duration of the experiment until the bubble has decreased to a size, when changes in its behaviour appear (around $2\,s$). However, other integration times, longer and shorter did not affect the overall observations qualitatively. In our analysis the forward and backward LCS carry the information of different time-intervals and need not be perpendicular to each other. We focus on the area below the bubble since we are interested in the fluid dynamics in the bubble wake. Also reflections of the bubble and the thinness of the fluid film at the bubble sides do not allow for a reliable LCS analysis in its direct vicinity. The entangled picture drawn by the LCS and the FTLE analysis resembles but also expands  the early classification in dynamically different parts such as vortex centers and stagnation point of the steady vortex pair below a rising bubble as for instance in\cite{Komasawa1980}. In contrast to the former research our results and other recent studies \cite{Falcone2018} deal with the realistic scenario of a temporally unsteady (and non-periodic) flow around rising bubbles. This is crucial for the influence of the hydrodynamics in context of mass transport since mixing with temporally changing manifolds is greatly improved in contrast to steady flow situations with constant streamlines.

We stress that we perform only a two-dimensional LCS and FTLE analysis since we only have velocity data from a two-dimensional plane cutting through our three-dimensional volume. It is known that this can lead to biased LCS-ridges and FTLE values. However, it was found that the two-dimensional FTLE analysis provides a good approximation whenever the strain in the third dimension is small\cite{Sulmann2013}. 
Here the corresponding strain is $S_{u,v}  = \sqrt{ (\delta u/\delta\phi)^{2}+  (\delta v/\delta\phi)^{2} }$ with $u,v$ being the in plane velocities and $\phi$ being the angle of rotation with regard to the center axis. Albeit a direct measurement of this strain was not feasible in our experiments we consider it to be small due to the strong rotational symmetry which is observed by eye for the two cases of lower Reynolds numbers ( $Re = 36$ and $Re = 156$ ). Only for the highest Reynolds number ($Re = 303$) we expect some deviations from a three-dimensional FTLE calculation since  we observe that eventually brighter fluid appears in a spot of dark fluid coming from a plane adjacent to the illuminated one. A clear hint of a non-negligible z-component of the velocity field. Therefore the analysis presented here needs to be treated with some caution in this case. 

\section{Results}\label{results}

\begin{figure}[p]
  \includegraphics[width=0.72\linewidth]{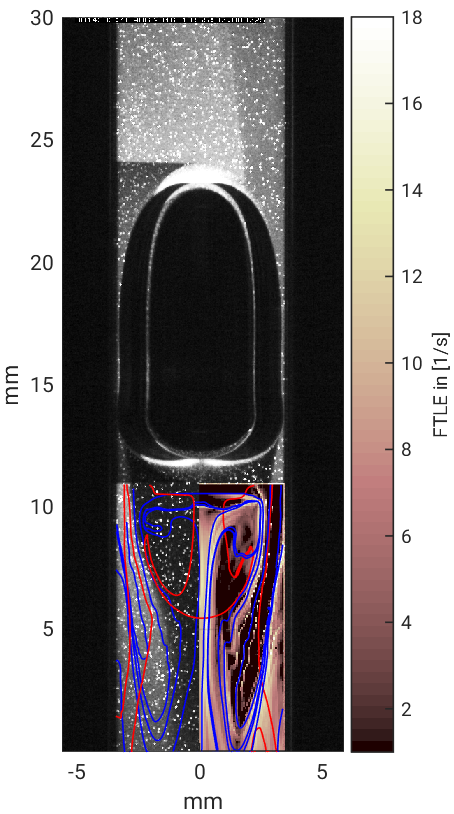}
  \captionof{figure}{LCS calculated from the velocity fields obtained from particle image velocimetry (PIV) data for a round channel diameter of $D = 7 \, \mathrm{mm}$. The corresponding rising velocity without counter flow would be $v_{rise} = 22 \, \mathrm{mm/s}$. The red lines denote  repelling LCS and blue lines the attracting LCS. On the right side below the bubble the backward FTLE-field $ \Lambda_{-}$ is shown in the background (color online). The liquid is an aqueous solution of fluorescein sodium salt which shows a decrease in fluorescence intensity for higher amounts of dissolved carbon dioxide gas released from the bubble. This decrease is visualized on the left side below the bubble using Laser Induced Fluorescence (LIF)\cite{Kastens2017}.}\label{fig:1} 
\end{figure}

Repelling material lines extracted by Lagrangian Analysis indicate where tracer will be separated in the future, while attracting material lines tell us where particles, previously far apart, have come together. This knowledge immediately gives us insight into the flow topology for the finite time $\tau$: imagine a tracer released at some point in the fluid but inside or outside (above or below) the red repelling LCS in Fig. \ref{fig:1}.  After a time $\tau$ the tracer released inside the red LCS will still largely remain within the vortical structures close to the bubble while tracer released below the red LCS will have been flushed away rapidly. This would also affect the local residence times of reactive molecules significantly. Especially the resident times for a product produced below the bubble is thus highly dependend on the local hydrodynamics and thus prone to be available for chemical reactions with longer timescales as desired. This can have undesired effects on overall yield and selectivity of the targeted reaction, especially if there are consecutive reactions of the product and the gas phase. The blue attracting LCS lines in Fig.\,\ref{fig:1} on the other hand show where particles have come together, i.e. where they were highly separated in backward time. The dark areas indicating high gas concentration by quenched fluorescence are largely enclosed by these blue LCS and also by high values of the backward FTLE. This is to be expected since the attracting LCS and the backward FTLE highlight the maximal stretching and transport barriers in backward time and thus the manifolds that have ordered the flow up to the time instant viewed. A similar reasoning explains why the forward LCS can not yet have an influence on the actual concentration distribution at the time instant viewed\cite{Nevins2017}. 

\begin{figure*}[t] 
\includegraphics [width=0.99\textwidth]{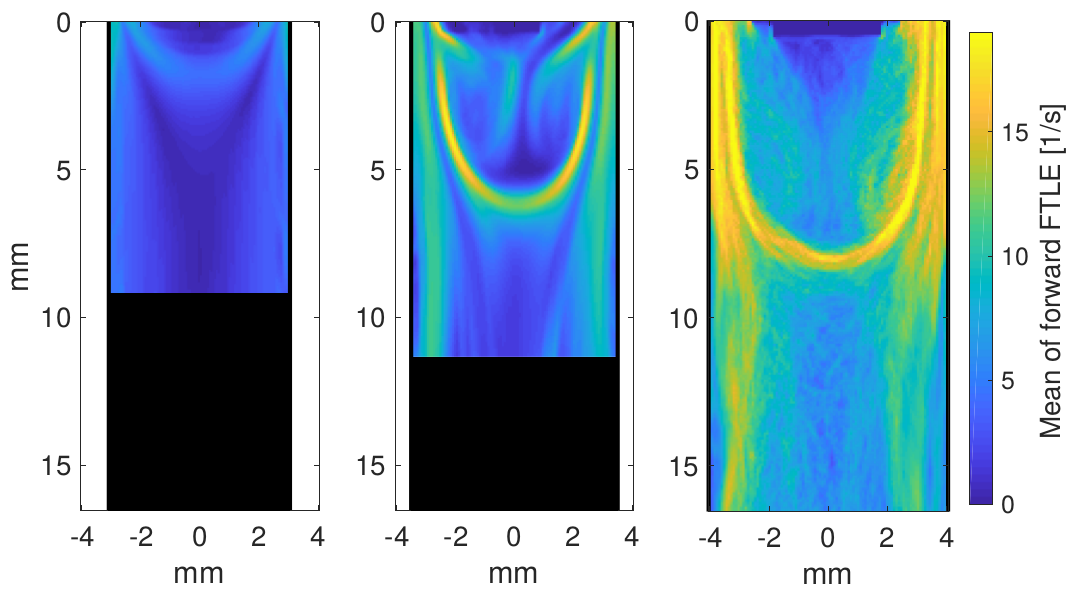}
  \captionof{figure}{Mean forward FTLE-field $\Lambda_{+} $ for three different flow regimes. Left to right: $D = 6\,mm$, $D = 7\,mm$, $D = 8\,mm$. High mean forward FTLE values show where the stretching is generally high as the FTLE can be viewed as the mean exponential stretching rate for particles that are initially close. Black color denotes areas where we have no information due to FOV restrictions in camera settings. }\label{fig:2}
\end{figure*}

 Further also the backward FTLE-field $\Lambda_{-}$ is shown in the lower right of Fig.\,\ref{fig:1}. The FTLE value at each point describes the maximum rate of deformation of an initially round infinitesimal circle for the finite time $\tau$ in forward or backward time respectively. Strong positive values of the forward FTLE (not shown) largely correspond to the red repelling LCS and separate the vortical structure in the direct bubble wake from the fluid that does not get entrained into the vortices and passes by with a velocity faster than the mean superficial velocity ahead of the bubble. Another red repelling LCS occurs regularly at the side border of the round channel where the fluid velocity is small. The overall picture of the flow topology derived from this example from time step $t_{i}$ is insensitive to a variation of the timestep and the integration time $\tau$ (not shown). We adress the persistence of LCS by taking the mean of the FTLE-fields for various timesteps. Figure \ref{fig:2} shows such a mean forward FTLE field of all instantaneous FTLE fields $t_{i} \in  \left[ \tau, 1000 \, \delta t-\tau  \right ] $ for all three flow regimes, where $\tau$ was again set to $120 \,\delta t$ and $\delta t = 1/600\,s$. It can be seen that in the intermediate and fastest flow regimes at $D = 7\,mm$ and $D = 8\,mm$ prominent horseshoe like structures appear in the mean forward FTLE fields which show the persistence of the coherent LCS structures found in the instantaneous flow fields. However, the structure in the case of the fastest flow (right figure) appears more fuzzy which can be understood by looking at the variance of the coherent structures (not shown). It is much higher in this case, therefore allowing particles to enter or leave the horseshoe region eventually at much faster relative timescales than for the intermediate case.
  
\begin{figure*}[t]
  \includegraphics[width=\textwidth]{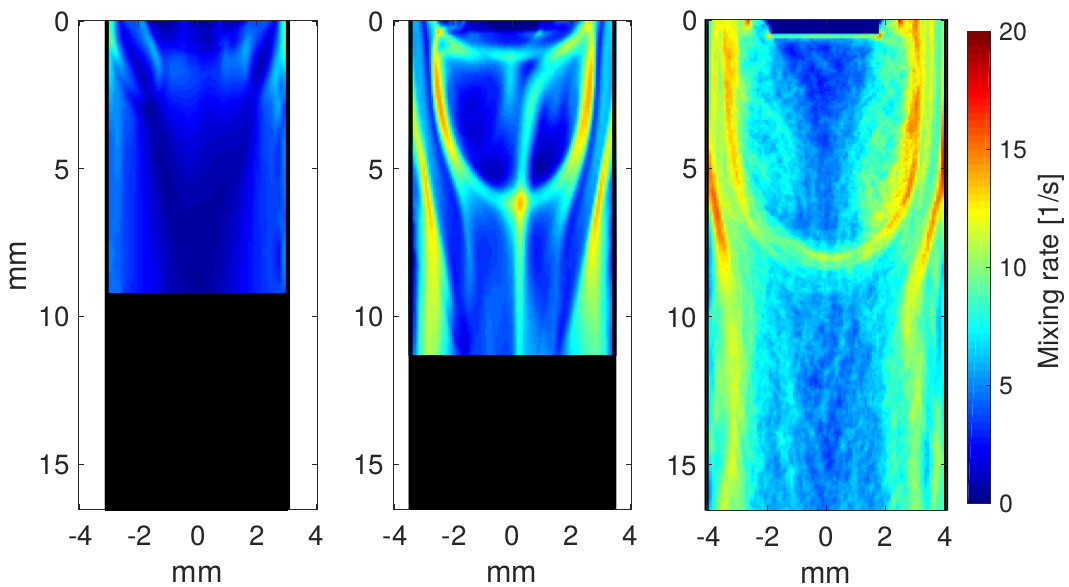}
   \captionof{figure}{The mean mixing rate as calculated from the FTLE fields via $M = 0.5 (<\Lambda_{+}> + <\Lambda_{-}>)$ is shown for three different flow regimes. Left to right: $D = 6\,mm$, $D = 7\,mm$, $D = 8\,mm$. High mean mixing rates denote areas where fluid is well mixed. }\label{fig:3}
\end{figure*}

Since the FTLE values are composed of deformation rates they can also be viewed as a measure of turbulent dispersion and thus of mixing. At a point in the round channel where both, the forward and the backward instantaneous FTLE values are high, the particles that pass trough this point at a distinct instant in time have come together from far apart and will also separate quickly during the time-interval considered ($\tau$). Thus the conditions for good mixing are met at this point and a tracer placed here at time $t_{i}$ will be in contact with fluid parcels of very different fate and origin. If this is true for the average over many  time instants $t_{i}$ the mean mixing at this point is high. Thus, to define a mean mixing rate we take the average of the mean forward and the mean backward FTLE values $M = 0.5 (<\Lambda_{+}> + <\Lambda_{-}>)$ derived from our FTLE analysis. Figure \ref{fig:3} shows this mean mixing rate for all three round channel diameters (i.e. Reynolds numbers). As in the previous figure there is no intense structure for the case of the slowest flow ($D = 6\,mm$). For the intermediate regime a well refined hyperbolic point appears where the mean mixing rate is very high. Also, at the outer edges of the horseshoe-like structure high mean mixing rates can be found. It is reasonable to assume that at these areas a chemical reaction relying on the dissolved gas and fresh chemicals from the passing liquid phase will reach maximum yields here if the timescales of the reactions are similar to the residence times. The high mixing values at the side walls of the round channel on the other hand cannot take part in such a reaction, since the dissolved gas is prevented from reaching the wall area by quickly accelerated solution which forms a transport barrier between the wall and the vortical structure (thin elongated blue areas of very low mixing close to zero). For the fastest regime on the contrary, a reasonable nonzero mixing rate is reached almost everywhere in the bubble wake which should soften the local concentration gradients and thus equalize the reaction kinetics at different locations in the wake.

\begin{figure*}
  \includegraphics[width=\textwidth]{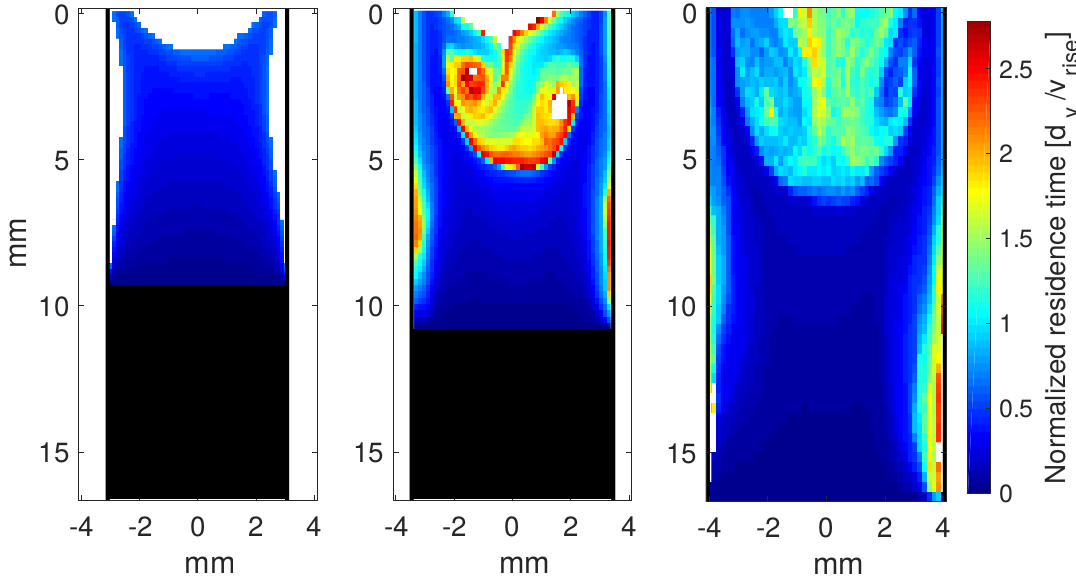}
    \captionof{figure}{The normalized mean local residence time distributions show a peak in residence times for intermediate Reynolds numbers. Left to right: $D = 6\,mm$, $D = 7\,mm$, $D = 8\,mm$.}\label{fig:4}
\end{figure*}

\begin{figure*}
  \includegraphics[width=\textwidth]{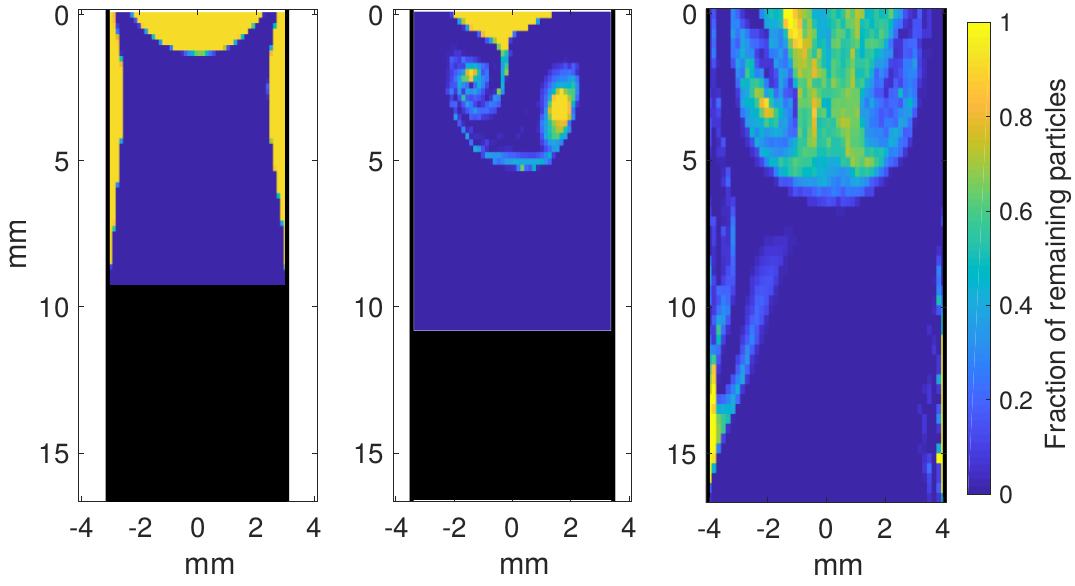}
    \captionof{figure}{The fraction of numerically applied particles at different release times $ t_{n} $ that do not leave the domain during the time of the experiment are shown as a colorcode. Note that especially within the vortex cores for intermediate Reynolds number no particle has left yet. Left to right: $D = 6\,mm$, $D = 7\,mm$, $D = 8\,mm$.}\label{fig:5}
\end{figure*}

However, each chemical reaction has a typical characteristic timescale $\tau_{reac}$ and it is therefore desirable to analyze the fluid flow according to  residence times of passive tracers, $\tau_{res}$. Both timescales can be directly compared to each other. Their ratio $Da  = \tau_{res}/ \tau_{reac}$  is a Damköhler number for chemical reactions\cite{Neufeld2004, Neufeld2010,Shah2012, Ottino1989}, here it measures the local mean ratio of an advective timescale $\tau_{res}$ to the reaction timescale. Figure \ref{fig:4} depicts the mean local  residence time distribution normalized to the mean residence time of a particle crossing the field of view (colored area) in flow direction with the velocity of the rising of the bubble $v_{rise}$, thus, $t_{norm} = d_{y}/v_{rise}$, where $d_{y}$ denotes the vertical dimension of the field of view. The mean local residence time distributions are calculated by the repeated release  of numerical particles at different timesteps $t_ {n}$ on a rectangular grid spanning the whole field of view shown ($100 \times 100 = 10000 $ equidistant initial positions) and their forward in time advection. The timestamp when a particle leaves the field of view is then written to its initial position for each release timestep $ t_{n} $ and the mean over all realizations is calculated. The instantaneous local residence time maps from a single release timestep $t_ {n}$ (not shown) resemble the mean local residence time distributions, which shows that there is no strong temporal intermittency. When comparing the normalized residence time distributions for the three Reynolds numbers it becomes immediately obvious that a maximal value is reached for the intermediate regime where a coherent vortex is the dominant structure in the velocity fields. White color denotes areas where no particle has yet left the domain for all initial times $ t_{n}$ and the whole data set. This behaviour is also depicted in Fig.\ref{fig:5} where the percentage of the particles that have not yet left the field of view in the time of the experiment  at each initial position is shown. 

\begin{figure*}
  \includegraphics[width=\textwidth]{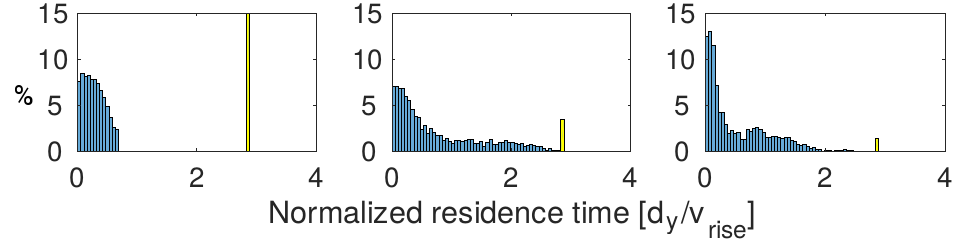}
    \captionof{figure}{The collapsed residence time distributions underpin the finding of a peak in residence times for the intermediate Reynolds number regime. Note that the yellow bar indicates the particles that have not yet left the domain. Left to right: $D = 6\,mm$, $D = 7\,mm$, $D = 8\,mm$.}\label{fig:6}
\end{figure*}

The collapsed residence time distribution of all particles that have left after the normalized time $t_{norm}$ (x-axis) in Fig.\,\ref{fig:6} undermines the observation of a maximum in the mean local residence time distribution in dependency of the Reynolds number.  The blue bars show the collapsed distribution of the mean residence times while the yellow bar signalizes the particles that have not yet left the domain. In a mere round channel flow without a bubble one would expect a linear decay in residence times and no tail for our initial tracer distribution. Here, in the intermediate regime at  $D = 7\,mm$, the effect of the wake of the bubble causes the collapsed residence time distribution to exhibit a significant tail which reaches values of  $t_{norm} = 3$ and higher which is significantly longer than for higher or lower Reynolds numbers. Further, there is a non zero valued gap in between the yellow and the blue bars for the intermediate case which hints that the tail of the distribution would be even more extended for longer experimental datasets (which were not available due to practical experimental restrictions). For higher or lower Reynolds numbers the tails are shorter and there is a range of zero values considerably before the yellow bar. At the lowest Reynolds number,  the particles that have not left the domain are particles placed directly below the bubble or at the side-walls where the velocities are so low, that within the accuracy of our PIV analysis they are zero. This might cause a bias towards higher sticking ratios for this Reynolds number. The instantaneous and the mean local residence time distributions show strong spatial inhomogeneities, especially for the intermediate Reynolds number regime. These large spatial gradients and the peak in residence times for a distinct Reynolds number have strong implications for the design of Taylor bubble reactors since it highlights the fact that the coherent structures that we detected can locally dominate the dynamics. Thus, faster mean superficial flow velocities are not trivially connected to shorter fluid flow timescales in the wake of the bubbles. 
This is of particular importance when the flow timescales are adjusted to reaction timescales in order to increase yield and production because in these particular cases when coherent structures form in the reactor one could obtain the opposite effect.

An interesting thought game is to imagine a competitive consecutive reaction as a model reaction in this Taylor bubble experiment $A+B \rightarrow P$ and $P+B \rightarrow S$, where $B$ is the gaseous phase such as was done in reference\cite{Roessler2001}. The reaction constants are $k_{1}$ and  $k_{2}$ respectively and $k_{1} > k_{2}$ which means that the first reaction is faster than the second one. Further, we assume that both characteristic reaction timescales in this model reaction are longer than the typical mean integral contact time the fluid parcel needs to pass by the bubble surface since otherwise the reaction would only take place at the interface of the bubble or its boundary layer. It would thus be dominated by the surface size and not by the details of the velocity fields in the bubble wake. Also we assume that $A$ exists in abundance everywhere (outside of the bubble) and the process occurs under isothermic conditions. With these settings for the regime of the low Reynolds number we would expect that basically no second product $S$ is formed since the source of $B$ is the bubble only such that dissolved gas gets solemnly transported straight downward and for the slower, consecutive reaction there is simply little $B$ left. If one wishes now to increase the overall production rate of $P$ it could be tempting to simply adjust the mean flow rate so the gas is more rapidly dissolved. However, our findings would contradict this viewpoint since the peak in residence times within the coherent vortex makes it more likely for the recirculating first product $P$ to find another molecule of $B$ right behind the bubble and thus pushes the generation of $S$. Thus we hypothesize that the yield of $P$ will not increase as expected and could even go down. Only for even faster flow rates when the coherent structures become unstable in time the overall yield of P would increase again since the strong recirculation ceases to exist and mixing is fast enough to disperse the gas quickly.  When comparing our reaction situation to the one discussed in Fig. 4 of article\cite{Roessler2001} we would find within the vortical structure for the intermediate Reynolds number case well mixed $A$ and $B$ while outside there would almost only be $A$. As in their Fig.1 much of $P$ will be produced but in contrast in our case there will always be fresh $B$ available due to the near source such that the consecutive product $S$ will finally become the dominant species. We want to point out that the above discussion is purely exemplary and should serve to obtain a conceptual understanding of the involved dynamics. It is clear that every realistic situation will be far more complex since the reaction kinetics depend on the details of the actual reactant concentrations. Also, other changes in properties of the reactives and the solvent could also affect the local transport\cite{Ottino1989,Neufeld2010}.
\vspace{3mm}

\section{Conclusion}
We have analyzed experimentally derived velocity fields from the wake of a Taylor bubble by taking on a Lagrangian perspective in order to extract important transport characteristics. This viewpoint is especially important when considering mass transport since it is inherently Lagrangian. The wake of the Taylor bubble exhibits different transport regimes in dependance of the Reynolds number. These were analyzed by looking at the LCS and FTLE, the mixing rates and the residence time distributions. FTLE and LCS analysis is a versatile tool for all flows that have a considerable time dependance such that knowing the streamlines of the mean field does not provide the full picture of the passive tracer dynamics\cite{Haller2000}. Most importantly, we find an increase of the residence time distribution for intermediate Reynolds numbers when coherent wake vortices form. This coherent vortices are known to exist also in freely rising bubbles and bubble columns\cite{Weiner2018,Timmermann2016,Komasawa1980} and might thus also play a role for global effects in those reactors. As analysed via LCS and FTLE the mixing is enhanced for increasing Reynolds numbers while the local gradients of mean mixing in finite time are highest for intermediate Reynolds numbers, again underpinning the importance of coherent structures in the flow. 
We discussed the effect of our findings on a hypothetical chemical reaction of competitive consecutive type and draw general ideas for the overall yield. However, we are aware that with a real chemical reaction also other effects may play a decisive role, such as the effect of the backward stretching rate on the local reaction dynamics such as shown to play a decisive role for excitable reactions\cite{Nevins2017}. To conclude we expect our findings to be of crucial interest for all reactive flows where coherent structures exist on comparable timescales as those of the reactions involved. Coherent structures can appear in many fluid flows, behind bubbles or wall edges (deadzones), static mixers and other flow obstacles.
Certainly, the details of the interaction of the coherent structures and the chemical reaction need to be investigated in more detail in the future and an experiment performing PIV and LIF simultaneously in a competitive consecutive reaction would be desirable. 

\acknowledgements{
The authors gratefully acknowledge the financial support provided by the German Research Foundation (DFG) within the Priority Program “Reactive Bubbly Flows”, SPP 1740 (SCHL 617/ 12-2 and HE 5480/10-2, http://www.dfg-spp1740.de/).}




\end{document}